\documentclass[aps,twocolumn,prb]{revtex4}
\usepackage[dvips]{graphicx}
\usepackage{bm}
\usepackage{amsmath}

\newcommand{\ab}{{\alpha\beta}}
\newcommand{\dab}{{\delta_\ab}}
\newcommand{\dij}{{\delta_{ij}}}
\newcommand{\rij}{{\bm{r}_{ij}}}
\newcommand{\rijn}{{\bm{r}_{ij}+\bm{n}}}
\newcommand{\kG}{{\bm{k}+\bm{G}}}
\newcommand{\Gunit}{{\frac{2\pi}{a_0}}}
\newcommand{\eInf}{{\epsilon_\infty}}
\newcommand{\La}{{L_\alpha}}
\DeclareMathOperator{\erfc}{erfc}

\begin{document}

\title{Ferroics: magnetic-compass lattice and \\
  optical phonon dispersions of dipolar crystals}

\author{Takeshi Nishimatsu$^{1}$}
\author{Umesh V. Waghmare$^{2}$}
\author{Yoshiyuki Kawazoe$^{1}$}
\author{Benjamin Burton$^{3}$}
\author{Kazutaka Nagao$^{4}$}
\author{Yoshihiko Saito$^{5}$}

\affiliation{
$^{1}$Institute for Materials Research (IMR), Tohoku University, Sendai 980-8577, Japan\\
$^{2}$Theoretical Sciences Unit,
Jawaharlal Nehru Centre for Advanced Scientific Research (JNCASR),
Jakkur, Bangalore, 560 064, India\\
$^{3}$National Institute of Standards and Technology (NIST),
100 Bureau Drive STOP 8520, Gaithersburg, MD 20899-8520\\
$^{4}$Research Institute of Electrical Communication (RIEC),
Tohoku University, Sendai 980-8577, JAPAN\\
$^{5}$Osaka Science Museum,
4-2-1 Nakanoshima, Kita-ku, Osaka 530-0005, Japan\\
}

\begin{abstract}
We report a simple safe and attractive pedagogic demonstration
with magnetic compasses that facilitates an intuitive understanding 
of the concept that ferromagnetism and ferroelectricity do
{\em not} result from dipole-dipole interactions alone.
Phonon dispersion relations were calculated for
a 3-dimensional simple-cubic dipole crystal
and a 2-dimensional square-lattice dipole crystal.
The latter calculation corresponds to the compass demonstration
that confirm the antiferro ground state structure.
A sum rule for the three optical phonon frequencies is discussed.
A mathematical non-analyticity in the longitudinal optical phonon are also illustrated.
\end{abstract}

\date{\today}


\maketitle
\section{Introduction}
Ferromagnetism and Ferroelectricity in crystals have occupied minds of
physicists and chemists for a long time.
Both are phenomena that involve temperature dependent
symmetry breaking.
Ferromagnetic and ferroelectric states
have parallel alignments of magnetic and electric dipoles,
respectively, in all unit cells of the crystal. Other possible 
orderings, e.g. antiparallel alignments of dipoles in 
neighboring unit cells, correspond to antiferro states.
Ferromagnetic and ferroelectric (or antiferro) states
having reduced symmetries are stable only up to a Curie (or N\'eel) temperature,
above which the crystal typically transforms to higher-symmetry paramagnetic or
paraelectric states.
Energetic stability of the low-temperature state
reflects the resolution of competition between various interactions. 
Ferroelectricity originates from a delicate balance between 
short-range bonding and long-range dipole-dipole interactions.
In magnets, dipole-dipole interactions are
relatively weak, and stability is governed mostly by short-range interactions,
such as direct exchange, super exchange, double exchange, etc.
Dipole-dipole interactions in magnets are important, however, in 
the energetics of domain formation. 

Dipole-dipole interactions vary as $\frac{1}{r^3}$, where $r$ is dipole-dipole
separation, and they are anisotropic. It is not straight-forward for students 
to intuitively understand the role of dipole-dipole interactions in the stability of 
macroscopically ordered states, and that ferromagnetic and ferroelectric states are
{\em not} the results of dipole-dipole interactions between atoms or sites.
Here, we introduce an attractive demonstration
with magnetic compasses which facilitates an intuitive understanding 
(Sec.~\ref{sec:demonstration}).
In Sec.~\ref{sec:phonon},
we also introduce easy analytic and computational calculations
of the phonon dispersion relations of dipole crystals,
which: 1) demonstrates that the dipole-dipole interactions
yield an antiferro ground state; 2) yields a simple picture of LO-TO 
splitting; 3) reveals a mathematical non-analyticity at the $\Gamma$ point
in the phonon spectrum that arises from infinite range interactions in crystals
(this last point is not readily appreciated from standard text book treatments). 
Finally, we summarize in Sec~.\ref{sec:summary}. 

\section{Magnetic compass demonstration}
\label{sec:demonstration}
\begin{figure}[t]
  \centering
    \includegraphics[width=60mm]{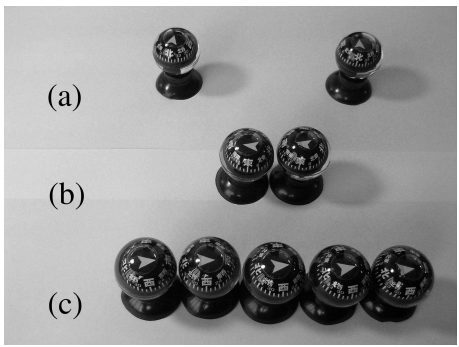}
    \caption{(a) Two car windshield magnetic compasses 
      on a desk, sufficiently far apart that both point north.
      (b) Two compasses sufficiently close together that the dipole-dipole 
          interaction dominates.
      (c) Five compasses in a row.}
    \label{fig:line}
\end{figure}

\begin{figure}[tb]
  \centering
    \includegraphics[width=60mm]{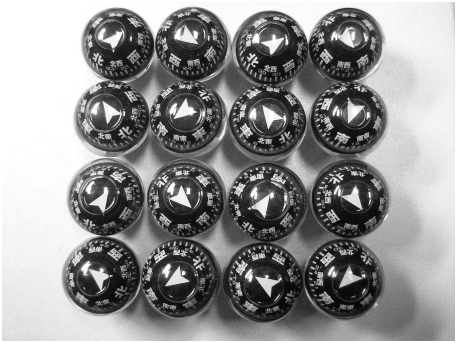}
    \caption{Magnetic compasses in a $4 \times 4$ square lattice.}
    \label{fig:square}
\end{figure}

If one puts two magnetic compasses on a desk well separated from each other,
both point north as pictured in Fig.~\ref{fig:line}(a).
As one brings them closer together,
their dipole-dipole interaction dominates and below some critical
distance, the influence of geomagnetism can be neglected.
The compass needles align along the line between them,
like $<\!{\rm S\,N}\!\!><\!{\rm S\,N}\!\!>$.
As in Fig.~\ref{fig:line}(c), if one puts three, four, or more compasses in a row,
one also finds that they align according to 
$<\!{\rm S\,N}\!\!><\!{\rm S\,N}\!><\!{\rm S\,N}\!>\cdots$.
If, however, one arranges the compasses on a square lattice,
they order in a stable canted antiferro structure,
which have zero total dipole moment, as shown in Fig.~\ref{fig:square}.

We demonstrated this experiment to undergraduate students
and had them do it themselves. It generated a lot of 
interests in them and helped them
to intuitively understand that 
ferromagnetic and ferroelectric ground states are do {\em not} 
occur a result of dipole-dipole interactions alone.
One author, YS,
originally invented this demonstration\cite{Saito:Yasue:FrontierPerspectives:2001}
and displayed it in the Osaka  Science Museum.
It attracted children as well as adults
seeding interest in the origin of magnetism in crystals.~\cite{Saito:ButsuriKyouiku:2005}

This experiment is easier to perform when one uses
magnetic compasses for car windshields, such as those shown  
in Figs.~\ref{fig:line}~and~\ref{fig:square}.
Each compass has a transparent bubble-like plastic container filled with oil,
a suction cup for attaching it to a windshield, and
a ferrite permanent magnet arrow that floats freely in the oil.
Dimensions of the ferrite magnet are $8\times 6\times 3$~mm.
The $8\times 3$~mm arrow sides are magnetic poles with surface 
magnetic fields $\approx$~50~mT,
that decay to $1.5$~mT at the surface of the 
plastic container.~\cite{Saito:ButsuriKyouiku:2005}

\section{Phonon dispersion of dipolar crystals}
\label{sec:phonon}
A system of classical dipoles on an array of sites is equivalent
to a system of paired point charges $+Z^*$ and $-Z^*$ on the same sites overlappingly,
in which the charges are allowed to move oppositely and infinitesimally:
the dipole moment of the displaced charges is ${\bm \mu}_i=Z^* {\bm u}_i$,
where ${\bm u}_i$ is
the displacement (separation vector) of $i$th pair point charges $\pm Z^*$.
Along the displacement, the pair point charges
do not shift their center of mass from their original site.
The dipole-dipole interaction
energy depends quadratically on the dipole moments ${\bm \mu}_i$ and
is directly related to harmonic phonon dispersion of the system.
Using the analysis of phonon spectrum,
we show the stability of the antiferro structure exactly and analytically.

It should be mentioned that there is a difference in degrees of freedom between
displacement ${\bm u}_i$ and a magnetic compass.
The number of degrees of freedom of ${\bm u}_i$ is three,
though that of a magnetic compass is one, i.e. its rotation.
However,
as shown in
the analysis of phonon spectrum for a 2-dimensional square-lattice dipole crystal
in Sec.~\ref{sub:phonon2d},
there is a correspondence between
the structure of phonon which has the minimum energy and
the ground state of compass lattice.

\subsection{3-dimensional simple-cubic dipole crystal}
\label{sub:phonon3d}
Consider a 3-dimensional simple-cubic dipole crystal
with lattice constant $a_0$.
Dipoles 
are placed on lattice points and
{\em only} the dipole-dipole interactions are considered,
\begin{equation}
  \label{eq:r3}
  V(\bm{u}_i, \bm{u}_j; \rij)=
  \frac{Z^{*2}}{\eInf}\frac{\bm{u}_i\cdot\bm{u}_j-
                            3(\widehat\rij\cdot\bm{u}_i)(\widehat\rij\cdot\bm{u}_j)}
                           {r_{ij}^3}~,
\end{equation}
where: $Z^*$ is the Born effective charge;
$\eInf$ is the optical dielectric constant, refractive index squared;
$\bm{u}_i=\bm{u}(\bm{r}_i)$ and $\bm{u}_j=\bm{u}(\bm{r}_j)$ are
the displacements of dipoles $i$ and $j$;
$\rij=\bm{r}_i - \bm{r}_i$ is a vector between dipoles $i$ and $j$; and
a hat indicates that the vector has a unit direction, e.g. $\widehat{\bm{r}}=\bm{r}/|\bm{r}|$.
The dipoles $\{\bm{u}\}$ in a super cell
$\Omega_{\rm super} = L_xa_0\times L_ya_0 \times L_za_0 = N a_0^3$
under periodic boundary conditions are governed by the {\em phonon} Hamiltonian:
\begin{widetext}
\begin{equation}
  \label{eq:Hamiltonian}
  H(\{\bm{p}\},\{\bm{u}\}) = \sum_{i=1}^N \sum_\alpha \frac{\{p_\alpha(\bm{r}_i)\}^2}{2M^*}
  + \frac{1}{2}\sum_{i=1}^N \sum_\alpha \sum_{j=1}^N \sum_\beta
  u_\alpha(\bm{r}_i) \Phi_\ab(\rij) u_\beta(\bm{r}_j)~,
\end{equation}
\end{widetext}
where $\alpha$ and $\beta$ are independent Cartesian directions ($x$, $y$, and $z$),
the $\alpha$ component of $\bm{r}$ runs:
\begin{equation}
  \label{eq:r}
  r_\alpha = 0,\ a_0,\ 2 a_0,\ \cdots\ (L_\alpha-1)a_0~,
\end{equation}
$M^*$ is the effective mass for dipoles,
$\bm{p}(\bm{r}_i)=M^*\dot{\bm{u}}(\bm{r}_j)$
is the momentum for $\bm{u}(\bm{r}_i)$,
$\Phi_\ab(\rij)$ is the force constant matrix
\begin{equation}
  \label{eq:Phi}
  \Phi_\ab(\rij)
  = \frac{Z^{*2}}{\eInf}\sum_{\bm{n}}\!'
  \frac{\dab - 3(\widehat\rijn)_\alpha(\widehat\rijn)_\beta}{|\rij + \bm{n}|^3}~,
\end{equation}
$\bm{n}$ is the super-cell lattice vector:
\begin{equation}
  \label{eq:n}
  n_\alpha = \cdots,\ -2\La a_0,\ -\La a_0,\ 0,\ \La a_0,\ 2\La a_0,\ \cdots\ \ \ ,
\end{equation}
and $\sum'$ indicates that the summation does not include terms
for which $\rij=\bm{n}=0$.
In the 3-dimensional simple-cubic lattice,
direct evaluation of sum in eq.~(\ref{eq:Phi}) is difficult,
because summation of $1/r^3$ in eq.~(\ref{eq:r3}) converges slowly.
However, it can be evaluated by using the Ewald sum technique
\cite{AllenTildesleyComputerSimulationOfLiquids1990,KittelAppendixB,OhnoK1KawazoeAppendixC}
as (in $Z^*=\eInf=1$ unit)
\begin{widetext}
\begin{multline}
  \label{eq:Real:Space:Ewald}
  \Phi_\ab(\rij) =
  \sum_{\bm{n}}\!'
  \left[
    \dab B(|\rijn|) - (\widehat\rijn)_\alpha(\widehat\rijn)_\beta C_{rr}(|\rijn|)
  \right]\\
  + \sum_{\bm{G}}\sum_{\bm{k}}\!'' (\widehat\kG)_\alpha(\widehat\kG)_\beta \frac{4\pi}{\Omega_{\rm super}}
  \exp\!\left(-\frac{|\kG|^2}{4\kappa^2}\right)
  \cos\!\left[(\kG)\cdot\rij\right]
  - \dij\dab\frac{2}{3}\frac{2}{\sqrt{\pi}}\kappa^3~,
\end{multline}
\end{widetext}
where $\bm{G}$ is a reciprocal lattice vector, and
$\bm{k}$ is a reciprocal vector in a first Brillouin zone of the unit cell such as
\begin{equation}
  \label{eq:G}
  G_\alpha = \cdots,\ -2\Gunit,\ -\Gunit,\ 0,\ \Gunit,\ 2\Gunit,\ \cdots\ \ \ ,
\end{equation}
\begin{equation}
  \label{eq:k}
  k_\alpha = -\frac{\La-1}{2\La}\Gunit,\ \cdots,\ -\frac{1}{\La}\Gunit,\ 0,\ %
  \frac{1}{\La}\Gunit,\ \cdots,\ \frac{1}{2}\Gunit\ \ \ ,
\end{equation}
while the summation is not performed for the $\bm{G}=\bm{k}=0$ case,
as indicated in eq.~(\ref{eq:Real:Space:Ewald}) by $\sum''$.
The Ewald summation is performed with decay functions
\begin{equation}
  \label{eq:decayB}
  B(r) = \frac{\erfc(\kappa r)}{r^3}
         + \frac{2}{\sqrt{\pi}}\kappa\frac{\exp(-\kappa^2 r^2)}{r^2}~,
\end{equation}
\begin{equation}
  \label{eq:decayCrr}
  C_{rr}(r) = \frac{3\erfc(\kappa r)}{r^3}
         + \frac{2}{\sqrt{\pi}}\kappa(2\kappa^2+\frac{3}{r^2})\exp(-\kappa^2 r^2)~,
\end{equation}
where  $\kappa$ is a reciprocal decay length which one can determine arbitrarily.

\begin{figure}
  \centering
    \includegraphics[width=50mm,clip,angle=-90]{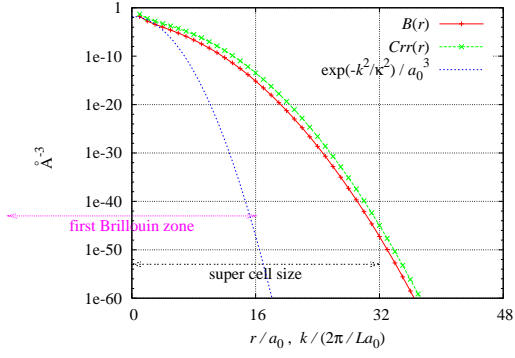}
    \caption{Decay functions
      $B(r)$ and $C_{rr}(r)$ in real space and
      $\exp(-k^2/\kappa^2)/a_0^3$ in reciprocal space
      for $a_0=3.94$~\AA, $L_x=L_y=L_z=32$, and $\kappa=0.078$~\AA$^{-1}$.
      These functions decay to less than $10^{-40}$~\AA$^{-3}$
      at the real or reciprocal zone boundaries.}
    \label{fig:DecayFunctions}
\end{figure}

It is well known that, by substituting
\begin{equation}
  \label{eq:u:IFT}
  u_\alpha(\bm{r}) = \frac{1}{N}
  \sum_{\bm{k}}\widetilde{u}_\alpha(\bm{k})\exp(i\bm{k}\cdot\bm{r})~,
\end{equation}
\begin{equation}
  \label{eq:p:IFT}
  p_\alpha(\bm{r}) = \frac{1}{N}
  \sum_{\bm{k}}\widetilde{p}_\alpha(\bm{k})\exp(i\bm{k}\cdot\bm{r})~,
\end{equation}
and
\begin{equation}
  \label{eq:Phi:IFT}
  \Phi_\ab(\bm{r}) = \frac{1}{N}
  \sum_{\bm{k}}\widetilde\Phi_\ab(\bm{k})\exp(i\bm{k}\cdot\bm{r})~,
\end{equation}
into eq.~(\ref{eq:Hamiltonian}),
the phonon Hamiltonian can be decomposed\cite{KittelAppendixC} into the
Hamiltonian which is {\em local} in $\bm{k}$-space,
\begin{equation}
  \label{eq:Hk}
  H_{\bm{k},\alpha} =\frac{\widetilde{p}^*_\alpha\!(\bm{k})
    \widetilde{p}_\alpha\!(\bm{k})}{2M^*} + 
  \frac{1}{2} \sum_\beta
  \widetilde{u}_\alpha^*(\bm{k}) \widetilde\Phi_\ab(\bm{k}) \widetilde{u}_\beta(\bm{k}),
\end{equation}
where
$\widetilde{u}_\alpha(\bm{k})$,
$\widetilde{p}_\alpha(\bm{k})$, and
$\widetilde\Phi_\ab (\bm{k})$ are Fourier coefficients,
\begin{equation}
  \label{eq:u:FT}
  \widetilde{u}_\alpha(\bm{k}) = \sum_{\bm{r}}u_\alpha(\bm{r})\exp(-i\bm{k}\cdot\bm{r})~,
\end{equation}
\begin{equation}
  \label{eq:p:FT}
  \widetilde{p}_\alpha(\bm{k}) = \sum_{\bm{r}}p_\alpha(\bm{r})\exp(-i\bm{k}\cdot\bm{r})~,
\end{equation}
and
\begin{equation}
  \label{eq:Phi:FT}
  \widetilde\Phi_\ab(\bm{k}) = \sum_{\bm{r}}\Phi_\ab(\bm{r})\exp(-i\bm{k}\cdot\bm{r})~.
\end{equation}
Here, $\widetilde\Phi_\ab(\bm{k})$ is a real function,
because $\Phi_\ab(\bm{r})$ is a real even function
in the simple cubic geometry.
The second term of eq.~(\ref{eq:Real:Space:Ewald}) is transformed as
\begin{equation}
  \label{eq:FT2}
  \widetilde\Phi_\ab^2(\bm{k})=
    \sum_{\bm{G}} (\widehat\kG)_\alpha(\widehat\kG)_\beta
    \frac{4\pi}{a_0^3}
    \exp\!\left(-\frac{|\kG|^2}{4\kappa^2}\right)~,
\end{equation}
and terms in the summation of eq.~(\ref{eq:FT2}) converge rapidly as a function of $\kG$.
Therefore, with optimal $\kappa$ and fast Fourier transformation (FFT)
for the first and third terms of eq.~(\ref{eq:Real:Space:Ewald}),
$\widetilde\Phi_\ab(\bm{k})$ can be evaluated by summing
only $n_\alpha = -L_\alpha a_0,\ 0$ terms in real space and a $\bm{G}=0$ term in reciprocal space.
For example, if $a_0=3.94$~\AA and $L_x=L_y=L_z=32$, 
$\kappa=0.078$~\AA$^{-1}$ gives sufficient accuracy in $\widetilde\Phi_\ab(\bm{k})$
for double precision computations, as illustrated in Fig.~\ref{fig:DecayFunctions}.

\begin{figure}
  \centering
    \includegraphics[width=75mm]{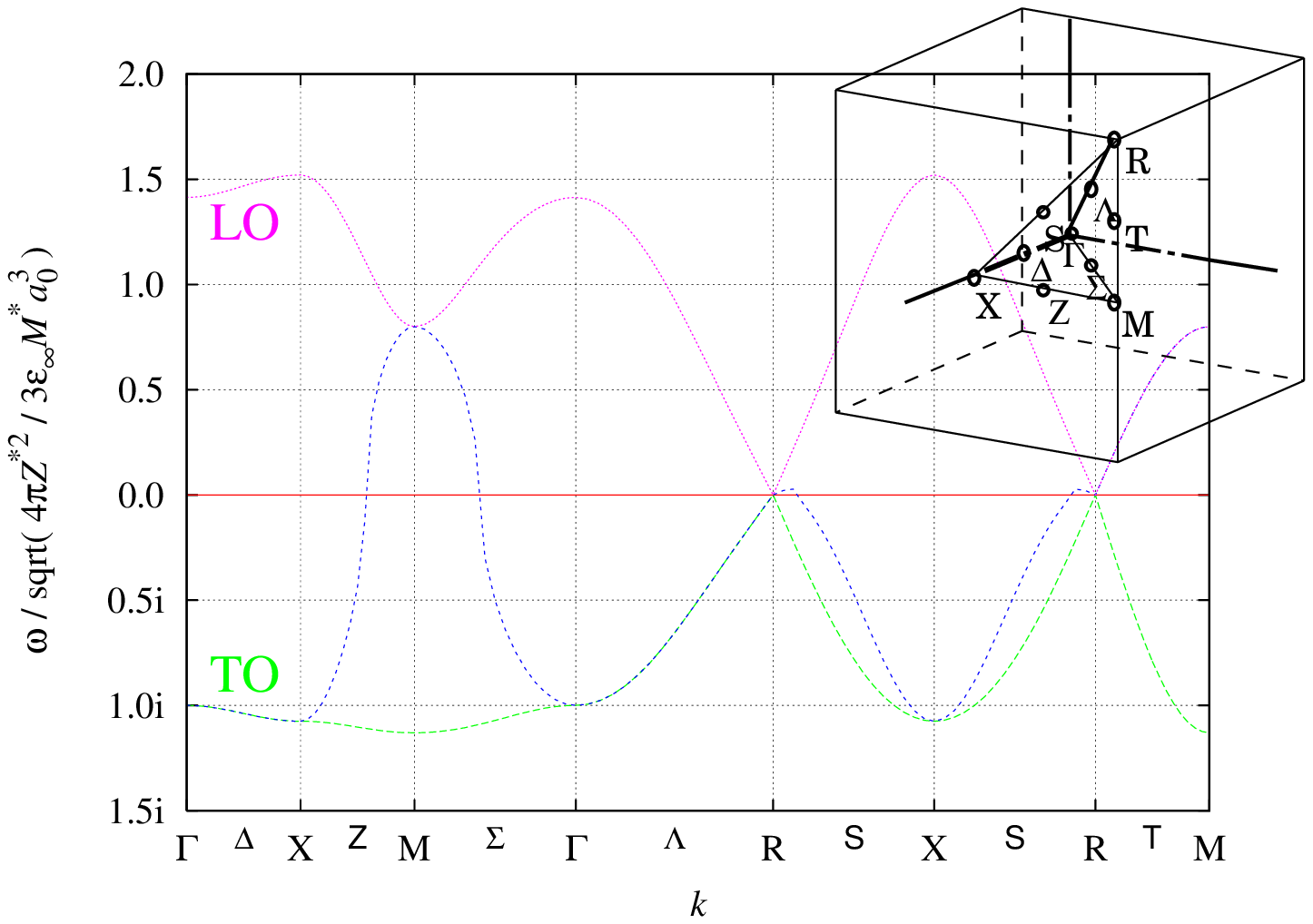}
    \caption{Phonon dispersion relations for the simple-cubic dipole crystal
      along axes of high crystal symmetry in the first Brillouin zone
      calculated with a $L_x\times L_y \times L_z = 32\times 32\times 32$ super-cell.
      Imaginary frequencies are plotted as negative values.
    }
    \label{fig:dispersion3}
\end{figure}

Three eigenvalues $\omega_1^2 \leq \omega_2^2 \leq \omega_3^2$ of
the $3\times3$ matrix $\widetilde\Phi_\ab(\bm{k})/M^*$
give longitudinal-optical (LO) and transverse-optical (TO)
phonon\footnote{Atomic-displacive waves in crystals accompanied by polarization
are called optical phonons, because they absorb lights in crystals.
The longitudinal-optical (LO) phonon has its atomic-displacive amplitude parallel to its wavevector.
The transverse-optical   (TO) phonon has that of perpendicular to its wavevector.}
dispersion relations of $\omega$ versus $\bm{k}$
as plotted in Fig.~\ref{fig:dispersion3},
where $M^*$ is the effective mass of a dipole.
There is a sum rule,
\begin{equation}
  \label{eq:sum:rule}
  \omega_1^2 + \omega_2^2 + \omega_3^2 = 0~,
\end{equation}
because
the trace of the matrix $\Phi_\ab(\rij)$ in eq.~(\ref{eq:Phi})
is zero for all $\rij$
and consequently the trace of $\widetilde\Phi_\ab(\bm{k})$ also equal to zero for all $\bm{k}$.

\begin{figure}
  \centering
    \includegraphics[width=60mm]{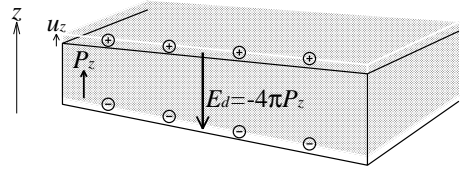}
    \caption{Schematic illustration of longitudinal oscillation of a thin slab or film of
      a dielectrics.}
    \label{fig:oscillation}
\end{figure}

In the long wavelength limit, $k\!\to\! 0+$
(with $L_\alpha\!\to\!\infty$ in the present super-cell calculation),
$\omega_1 = \omega_2$ converge
to            $\sqrt{-\frac{1}{3}\frac{4 \pi Z^{*2}}{\epsilon_\infty M^* a_0^3}}$ and
$\omega_3$ to $\sqrt{ \frac{2}{3}\frac{4 \pi Z^{*2}}{\epsilon_\infty M^* a_0^3}}$.
This frequency difference is called as the LO-TO splitting
which has a non-analyticity (discontinuity)
in a LO-branch
at the $\Gamma$ point, ${\bm k=(000)}$.
The LO-TO splitting is analogous to the plasma-oscillation\cite{KittelChap10} and
corresponds to the longitudinal oscillations of dielectrics
which obeys an equation of motion,
\begin{equation}
  \label{eq:oscillation}
  M^*\frac{d^2 u_z}{dt^2} = \frac{E_d}{\epsilon_\infty}Z^*
                         =-\frac{4\pi P_z}{\epsilon_\infty}Z^*
                         =-\frac{4\pi Z^{*2}}{\epsilon_\infty a_0^3}u_z~,
\end{equation}
where $E_d=-4\pi P_z=-4\pi Z^* u_z / a_0^3$ is the depolarization field 
that is induced by the surface charges that are caused by the polarization $P_z$,
as illustrated in Fig.~\ref{fig:oscillation}.
Along the $\Delta$ axis,
the point group of $\bm{k}$ is $C_{4v}$.
Consequently,
two TO phonon frequencies, $\omega_1 = \omega_2$, are degenerate,
non-degenerate $\omega_3$ has to be the singly degenerate LO phonon, and
their ratio has to be $\frac{\omega_3}{\omega_1} = \frac{\sqrt{2}}{i}$
to satisfy the sum rule (\ref{eq:sum:rule}).
The singly degenerate TO phonon at the M point has
the structure of oppositely $z$-polarized rods
in a checker-board arrangement on an $xy$ plane.
This antiferro structure has minimum energy.
At the R point, oppositely directed dipoles align in NaCl structure
and its three optical phonon frequencies become zero,
$\omega_1 = \omega_2 = \omega_3 = 0$,
because summing eq.~(\ref{eq:r3}) within $n$-th neighbors gives zero.
For instance, within 6 nearest neighbors,
\begin{equation}
  \label{eq:nearest:neighbors}
  \sum_{i\in{\rm (6~nearest~neighbors)}}V(\bm{u}, -\bm{u}; \bm{r}_{0i}) = 0~,
\end{equation}
and also within 12 second nearest neighbors,
\begin{equation}
  \label{eq:second:nearest:neighbors}
  \sum_{i\in{\rm (12~second~nearest~neighbors)}}V(\bm{u}, \bm{u}; \bm{r}_{0i}) = 0~.
\end{equation}

\subsection{2-dimensional square-lattice dipole crystal}
\label{sub:phonon2d}
\begin{figure}
  \centering
    \includegraphics[height=70mm,clip,angle=-90]{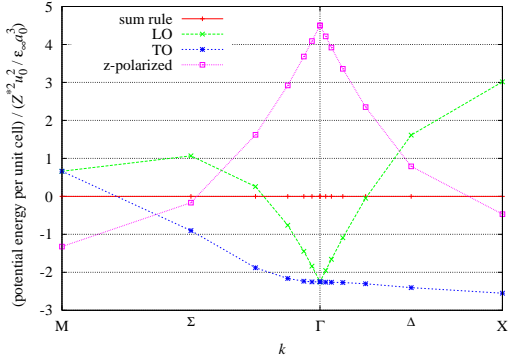}
    \caption{Energies of phonon $\bm{u}(\bm{r}_i) = \bm{u}_0\exp(i\bm{k}\cdot\bm{r}_i)$
      for the 2-dimensional square-lattice dipole crystal,
      as functions of wave vector $\bm{k}$ are plotted,
      along axes M --$\Sigma$-- $\Gamma$ --$\Delta$-- X in the first Brillouin zone.
      $N = L_x \times L_y = 64 \times 64$ is used for this calculation.
      This dispersion corresponds to
      TABLE II in Ref.~\cite{DeBell:M:W:RevModPhys:72:p225-257:2000}
    }
    \label{fig:energy2dim}
\end{figure}

\begin{figure}
  \centering
    \includegraphics[width=70mm]{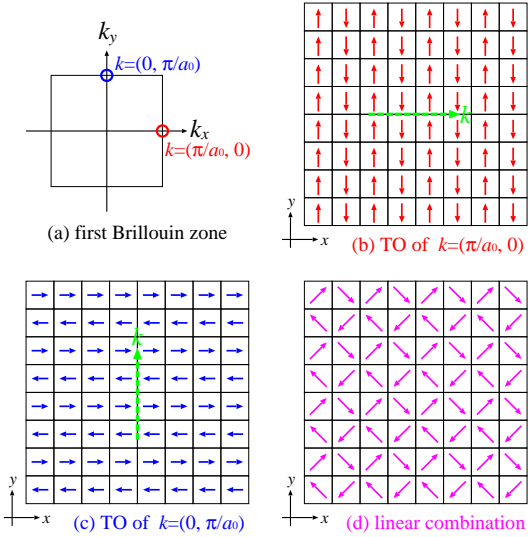}
    \caption{(a) First Brillouin zone of the 2-dimensional square-lattice crystal.
      Two X points $\bm{k}=(\pi/a_0, 0)$ and $\bm{k}=(0, \pi/a_0)$ are indicated.
      (b) Transverse-optical (TO) phonon of $\bm{k}=(\pi/a_0, 0)$ which is lowest in energy,
      i.e. $\bm{u}(\bm{r}_i) = (0,1,0)\exp[i\frac{\pi}{a_0}(\bm{r}_i)_x]$,
      is illustrated with arrows at each site.
      The direction of wavevector $\bm{k}$ is indicated with dashed arrow.
      (c) Another lowest-in-energy TO phonon of $\bm{k}=(0, \pi/a_0)$,
      i.e. $\bm{u}(\bm{r}_i) = (-1,0,0)\exp[i\frac{\pi}{a_0}(\bm{r}_i)_y]$.
      (b) A canted antiferro structure, which is a linear combination of (b) and (c),
      i.e. $\bm{u}(\bm{r}_i) = (0,1,0)\exp[i\frac{\pi}{a_0}(\bm{r}_i)_x]
                             +(-1,0,0)\exp[i\frac{\pi}{a_0}(\bm{r}_i)_y]$.
      }
    \label{fig:kxky}
\end{figure}

\begin{figure}
  \centering
    \includegraphics[width=70mm]{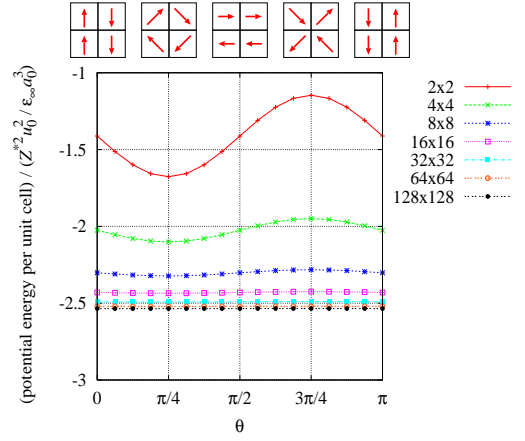}
    \caption{Comparison of energies as functions of
      the phase $\theta$ of two coefficients in eq.~(\ref{eq:theta})
      for finite-size 2-dimentional square-lattice dipole crystals
      from $2 \times 2$ to $128 \times 128$.
      $2 \times 2$ configurations of dipoles
      are also shown in above panels
      for $\theta = 0, \frac{\pi}{4}, \frac{\pi}{2}, \frac{3\pi}{4},$
      and $\pi$ (from left to right).
      The canted antiferro structure ($\theta=\frac{\pi}{4}$) gives
      minimum energies for all system sizes.
      Energy range becomes narrower and
      the energy converges to the value of infinite system ($\approx-2.549$),
      as the system size gets larger.}
    \label{fig:theta}
\end{figure}

The 2-dimensional square-lattice dipole crystal,
which has $N = L_x \times L_y$ dipoles in a flat periodic super cell,
is also governed by the Hamiltonian (\ref{eq:Hamiltonian}) and
the force constant matrix (\ref{eq:Phi}),
but there are no dipoles for $r_z\neq 0$ and $n_z\neq 0$.
In the 2-dimensional case,
summation in eq.~(\ref{eq:Phi}) can be performed in real space.
Calculated potential energies of phonons
$\bm{u}(\bm{r}_i) = \bm{u}_0\exp(i\bm{k}\cdot\bm{r}_i)$
having wave vectors $\bm{k}$ and
constant amplitudes $\bm{u}_0$ are shown in Fig.~\ref{fig:energy2dim}.
Eigenmodes of polarization are
$\bm{u}_0=(1,0,0)$, $\bm{u}_0=(0, 1,0)$, and $\bm{u}_0=(0,0,1)$ for $\bm{k}$ along $\Delta$ axis and
$\bm{u}_0=(1,1,0)$, $\bm{u}_0=(1,-1,0)$, and $\bm{u}_0=(0,0,1)$ for $\bm{k}$ along $\Sigma$ axis.
The TO phonon energy minimum is at the X point.
There are {\em two} X points,
$\bm{k}=(\pi/a_0, 0)$ and $\bm{k}=(0, \pi/a_0)$,
in the first Brillouin zone of the 2-dimensional square-lattice crystal
as depicted in Fig.~\ref{fig:kxky}(a).
Corresponding two antiferro structures,
$\bm{u}(\bm{r}_i) =  (0,1,0)\exp[i\frac{\pi}{a_0}(\bm{r}_i)_x]$ and
$\bm{u}(\bm{r}_i) = (-1,0,0)\exp[i\frac{\pi}{a_0}(\bm{r}_i)_y]$,
are illustrated in Fig.~\ref{fig:kxky}(b) and (c), respectively,
in which oppositely polarized rows of dipoles are arranged alternately.
A linear combination of these degenerate two TO phonons at different $\bm{k}$-points,
$\bm{u}(\bm{r}_i) = (0,1,0)\exp[i\frac{\pi}{a_0}(\bm{r}_i)_x]
                  +(-1,0,0)\exp[i\frac{\pi}{a_0}(\bm{r}_i)_y]$,
has a canted antiferro structure illustrated in Fig.~\ref{fig:kxky}(d).
In the 2-dimentional infinite-size system,
any linear combinations of the two lowest-in-energy TO phonons
form a ``continuously degenerate manifold of
antiferro ground states''\cite{DeBell:M:B:W:PRB:55:p15108-15118:1997}.
In finite-size systems, however, the continuous degeneracy breaks.
The linear combination can be generally expressed
with two coefficients $\cos\theta$ and $\sin\theta$:
\begin{multline}
  \label{eq:theta}
  \bm{u}(\bm{r}_i) = (0,1,0)\cos\theta\exp[i\frac{\pi}{a_0}(\bm{r}_i)_x]\\
                   +(-1,0,0)\sin\theta\exp[i\frac{\pi}{a_0}(\bm{r}_i)_y]~,
\end{multline}
and, in the finite-size system,
its energy becomes a function of $L_x$, $L_y$, and $\theta$.
In $L_x \times L_y = 2^n \times 2^n$ ($n=$ 1, 2,~$\cdots$) system,
as shown in Fig.~\ref{fig:theta},
$\theta=0$ gives the simple antiferro structure (Fig.~\ref{fig:kxky}(b)),
$\theta=\frac{\pi}{4}$ gives the lowest-in-energy
canted antiferro structure (Fig.~\ref{fig:kxky}(d)),
and $\theta=\frac{3\pi}{4}$ the highest-in-energy structure.
Therefore, the canted antiferro structure is realized in
the square lattice of magnetic compasses
as described in Sec.~\ref{sec:demonstration}.
This is because the canted antiferro structure
minimizes leak flux outside the system.
On the other hand,
the highest-in-energy $\theta=\frac{3\pi}{4}$ structure
maximizes leak flux.
It can be also seen in Fig.~\ref{fig:theta} that,
as the system size gets larger and consequently the finite-size effect get smaller,
energy range becomes narrower and
the energy converges to the value of infinite system ($\approx-2.549$).

\begin{figure}
  \centering
  \includegraphics[width=60mm]{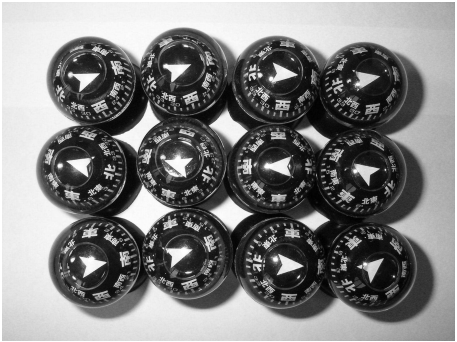}
  \caption{Magnetic compasses in a $4 \times 3$ square lattice.}
  \label{fig:4x3}
\end{figure}

It should be additionally mentioned that,
in $L_x \times L_y = 4 \times 3$ finite system
as shown in Fig.~\ref{fig:4x3} for example,
there may be another energetics, because it has odd $L_y$ and
the canted antiferro structure
cannot minimize leak flux outside the system.
Particularly in such small finite size systems,
sizes and shapes play important roles in the break of degeneracy.

\section{Summary}
\label{sec:summary}
We introduced an attractive demonstration
that uses magnetic compasses to facilitates an intuitive understanding of the fact
that ferromagnetic and ferroelectric orderings in crystals are
{\em not} the results of dipole-dipole interactions.
It is an easy and safe experiment that may be a good prologue to
lectures on ferromagnets and ferroelectrics.
Phonon dispersion of
3-dimensional simple-cubic and
2-dimensional square-lattice dipole crystals
are calculated to determine which antiferro structure is most stable,
when dipole-dipole interactions alone are considered. 
For the 1-dimensional chain of dipoles,
the $\Gamma$-mode with polarization along the chain
is lowest in energy.
For the 2-dimensional square-lattice dipole crystal,
the antiferro X-mode with transverse polarization
is lowest in energy.
For the 3-dimensional simple cubic dipole crystal,
the antiferro M-mode with a checker-board-arrangement of oppositely $z$-polarized rods
is lowest in energy.
The LO-TO splitting and a sum rule for the three optical phonon frequencies was
demonstrated in a simple way.
This is also a good example of application of the Ewald summation technique
commonly used in simulations of periodic systems of charged particles.

\section*{Acknowledgment}
We thank International Frontier Center for Advanced Materials (IFCAM) of IMR
who supported UVW to visit to Sendai and authors' collaborative study
in ferroelectrics.
We would like to extend our gratitude to
Professor Gyo Takeda (Emeritus Professor of University of Tokyo and Tohoku University) and
Professor Noboru Takigawa (Department of Physics, Tohoku University)
who showed TN the demonstration with magnetic compasses invented by YS.
We also extend our gratitude to
Professor Takayuki Hamaguchi (Nada Junior and Senior High School)
who gave suggestive photos and comments of the compass lattices to the authors.

\bibliographystyle{apsrev}
\bibliography{biblio/book,biblio/compasses}
\end{document}